\documentclass[a4paper]{jpconf}
\usepackage{graphicx}
\begin{document}
\title{Resolving the outer density profile of dark matter halo in Andromeda galaxy}

\author{Takanobu Kirihara, Yohei Miki and Masao Mori}

\address{University of Tsukuba, Tennodai 1-1-1, Tsukuba, Ibaraki, Japan}

\ead{kirihara@ccs.tsukuba.ac.jp}

\begin{abstract}
Large-scale faint structure detected by the recent observations in the halo of the Andromeda galaxy (M31) provides an attractive window to explore the structure of outer cold dark matter (CDM) halo in M31.
Using an $N$-body simulation of the interaction between an accreting satellite galaxy and M31, we investigate the mass density profile of the CDM halo.
We find the sufficient condition of the outer density profile of CDM halo in M31 to reproduce the Andromeda giant stream and the shells at the east and west sides of M31.
The result indicates that the density profile of the outer dark matter halo of M31 is a steeper than the prediction of the theory of the structure formation based on the CDM model.

\end{abstract}

\section{Introduction}
The cold dark matter (CDM) model, which is the standard paradigm for cosmological structure formation, predicts that galaxies have experienced many mergers with less massive galaxies and have grown larger.
Cosmological simulations of structure formation suggest that dark matter halos associated with galaxies have some universal density profile under an assumption of spherical symmetric halo model such as Navarro-Frenk-White (NFW) profile \cite{r14} or Fukushige-Makino-Moore (FMM) profile \cite{r4} \cite{r12}.
In these profiles, much controversy exists over the mass-density distribution of the inner dark matter halo.
With respect to the density distribution of the outermost region, however, the mass-density always decreases with the cube of the distance from the galactic center in these profiles.
That is to say the outer region of dark matter halo is the excellent laboratory to examine the prediction of CDM model.
However, it is extremely difficult to measure the mass distribution of the outer region of a galaxy because the stellar and/or gas density is too low to detect even by the latest instruments.
Consequently, observational verifications of the theoretical prediction are not yet enough so far.

Only recently, a deep and wide-field photometric survey of the Andromeda galaxy (M31), which is the nearest large galaxy, revealed that the outer region of M31 showed a wealth of substructures such as the Andromeda giant stream and the shells at the east and west sides of M31\cite{r7} \cite{r9}.
The Andromeda giant stream extends about $120$ kpc further away along the line of sight from M31 \cite{r11}, and 
the distribution of the radial velocity has already been observed \cite{r8}. 
These large-scale structures have been considered the evidence of a galaxy collision occurred about 1 Gyr ago \cite{r2} \cite{r13} \cite{r18}.
The $N$-body simulations in earlier studies examined the orbits if an infalling satellite dwarf galaxy which reproduce all of these spatial structures \cite{r2} \cite{r1} \cite{r3}.
In addition, Mori \& Rich \cite{r13} showed that the collision of M31 and the infalling satellite with the total mass ranging from $5\times10^8 M_{\odot}$ to $5\times10^9 M_{\odot}$ successfully reproduced the Andromeda giant stream and the shells by the $N$-body simulations.

Such a large-scale structure spreads far beyond the galactic disk and is suitable for the study of the actual density profile of the dark matter halo in M31.
Nevertheless, all of the earlier simulations always assumed that the mass density of the outer dark matter halo of M31 decreased with the cube of the distance from the galactic center in accordance with the prediction of the CDM model.
These situations motivate us to test the prediction of the CDM model on the density profile of the dark matter halo, using the $N$-body simulation for the formation of the Andromeda giant stream in the dark matter halo with the different density profiles.
In this paper, we introduce our numerical model in \S 2. In \S 3, we present simulation results, and a brief summary is given in \S 4.

\section{Numerical model}
In this study, we simulate the interaction between an accreting satellite dwarf galaxy and M31 using $N$-body simulations, which especially concentrated on the density profile of the dark matter halo in M31.
It should be noted that Mori \& Rich (2008) \cite{r13} studied the self-gravitating response of the disk, bulge, and dark matter halo of M31 to an accreting satellite and concluded that satellites less massive than $5\times10^9M_\odot$ had a negligible effect on the gravitational potential of M31.
Consequently, in this study, we treated M31 as the source of a fixed gravitational potential composed of a disk, a bulge, and a dark matter halo. 
For bulge and disk component, we adopt the model which reproduce the profile of surface brightness of M31 disk and bulge, rotation velocity curve of disk and velocity dispersion of bulge \cite{r2} \cite{r5}.

The NFW model that is widely accepted density profile of CDM halos was empirically derived from cosmological $N$-body simulation \cite{r14}. 
The resultant empirical profiles of density distribution is approximately fitted by
\begin{equation}
 \rho_{\mathrm{NFW}}(r)=\frac{\rho_{s}}{(r/r_{s})(1+r/r_{s})^2},
\end{equation}
where $r_{s}$ and $\rho_{s}$ are the scale radius and the scale density, respectively.
We focus on making a diagnosis of the density profile in the outer CDM halo using the Andromeda giant stream and the $N$-body experiments. In this purpose, we introduce the index $a$ in the equation of the density distribution as described below:
\begin{equation}
 \label{eq:eq1}
 \rho_{\mathrm{DMhalo}}(r)=\frac{\rho_{s,\;a}}{(r/r_{s,\;a})(1+r/r_{s,\;a})^a},
\end{equation}
where $r_{s,\;a}$ and $\rho_{s,\;a}$ are the scale radius and the scale density, respectively.
They are determined in such a way that the enclosed mass at $7.63$ kpc and $125$ kpc are identical for all models to reproduce the observed rotation velocity curve of M31.
We vary the parameter $a$ ranging from $1.5 < a < 4.0$ in this paper. In this case, NFW profile corresponds to $a=2$.

To investigate the dynamical response of the orbiting satellite, we modeled the satellite dwarf galaxy by a self-consistent $N$-body realization of stars under the influence of an external force provided by M31.
The satellite dwarf galaxy is adopted as a Plummer's sphere \cite{r17} with 49,152 particles.
The total mass is $2.2\times10^9 M_{\odot}$, and the scale radius is $1.03$ kpc.
The initial position vector and velocity vector for the standard coordinates centered on M31 are taken from Fardal et al. (2007) \cite{r2}.
Computation has been done on the FIRST cluster and Blade-GRAPE on FIRST in calculating self-gravity of satellite particles at Center for Computational Science (CCS), University of Tsukuba.\\

\section{Results}
The result of simulations indicates that the first pericentric passage occurred about $0.8$ Gyr ago, and the satellite collides almost head-on with M31. Then, the distribution of satellite particles is distorted and is spread out significantly. A large fraction of the satellite particles acquires a high velocity relative to the center of M31. This debris expands to a great distance, remaining collimated and giving rise to the Andromeda giant stream. 
Stellar particles that initially constituted the satellite start to form a clear shell structure at the north east area of M31. After the formation of the eastern shell, stars move to the west area and produce the western shell. 
Figure 1 shows the surface mass-density of the disrupted satellite galaxy at the present-day epoch. Each panel corresponds to the case for $a=2.0$ (upper left), the case for $a=2.5$ (upper right), the case for $a=3.0$ (lower left), and the case for $a=3.5$ (lower right). The positional coordinates $\xi$ and $\eta$ are the eastern and the northern direction on the sky, respectively. They are centered on the center of M31 and $1^{\circ}$ in angle corresponds to $13.6$ kpc.
Black circles show the edges of the observed shells taken from Table 1 of Fardal et al. (2007) \cite{r2}, and blue circles corresponds to the observed position of the Andromeda giant stream taken from Font et al. (2006) \cite{r3}.

\begin{figure}[h]
 \includegraphics[width=40pc]{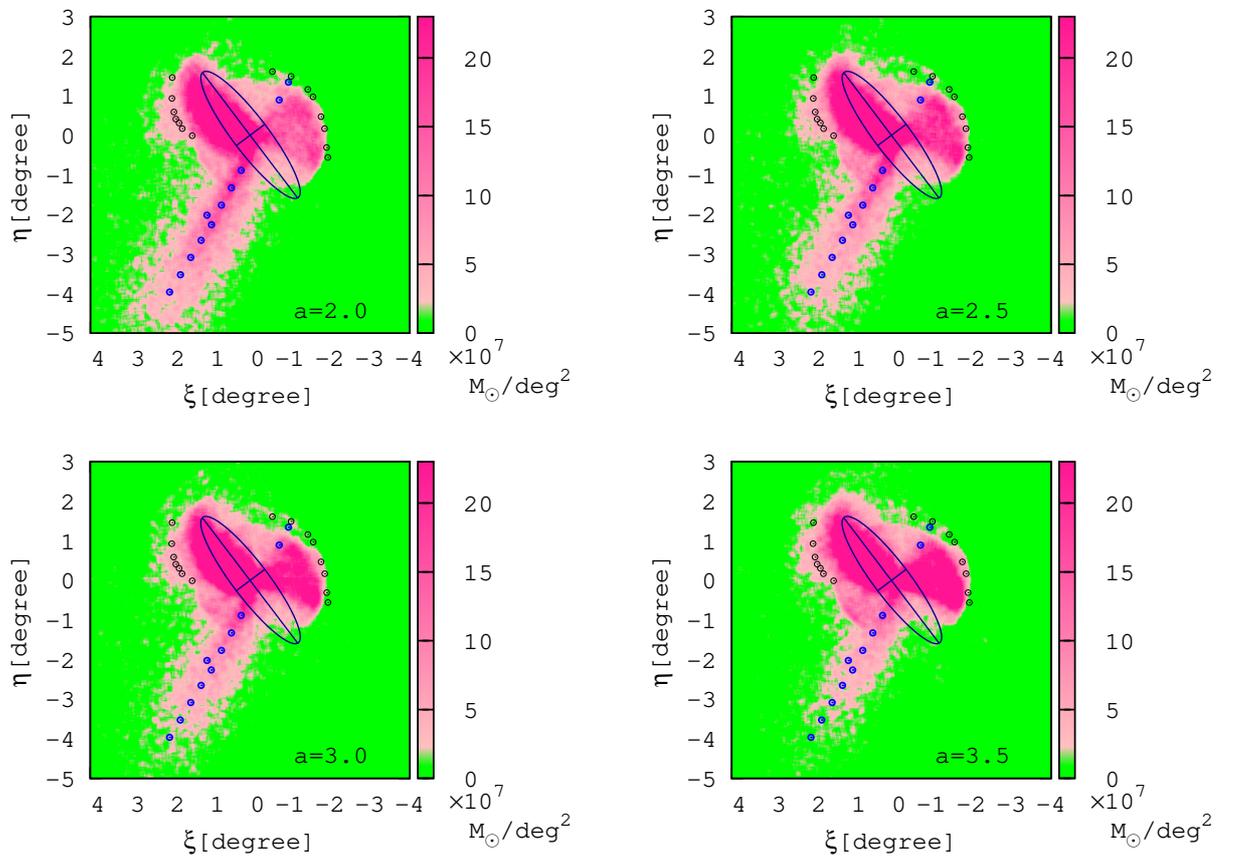}\hspace{2pc}%
 \begin{minipage}[b]{38pc}\caption{\label{label1}
Surface mass-density of the satellite galaxy at the present-day epoch. Each panel corresponds to the case for $a=2.0$ (upper left), the case for $a=2.5$ (upper right), the case for $a=3.0$ (lower left), and the case for $a=3.5$ (lower right).
The positional coordinates $\xi$ and $\eta$ are the eastern and the northern direction on the sky, respectively. They are centered on the center of M31 and $1^{\circ}$ in angle corresponds to $13.6$ kpc.
}
 \end{minipage}
\end{figure}
When the parameter $a$ becomes greater, the surface density of the Andromeda giant stream becomes lower.
This difference mainly comes from the difference of enclosed mass of dark matter halo around the Andromeda giant stream. The larger enclosed mass corresponds to the shorter free-fall time. In other words, the model with a large $a$ further accelerates the dynamical evolution of the whole system. 
Since stellar particles move quickly to the eastern shell after the formation of the Andromeda giant stream, the Andromeda giant stream has low surface density in case of a large $a$.

\begin{figure}[h]
 \begin{center}
  \includegraphics[width=38pc]{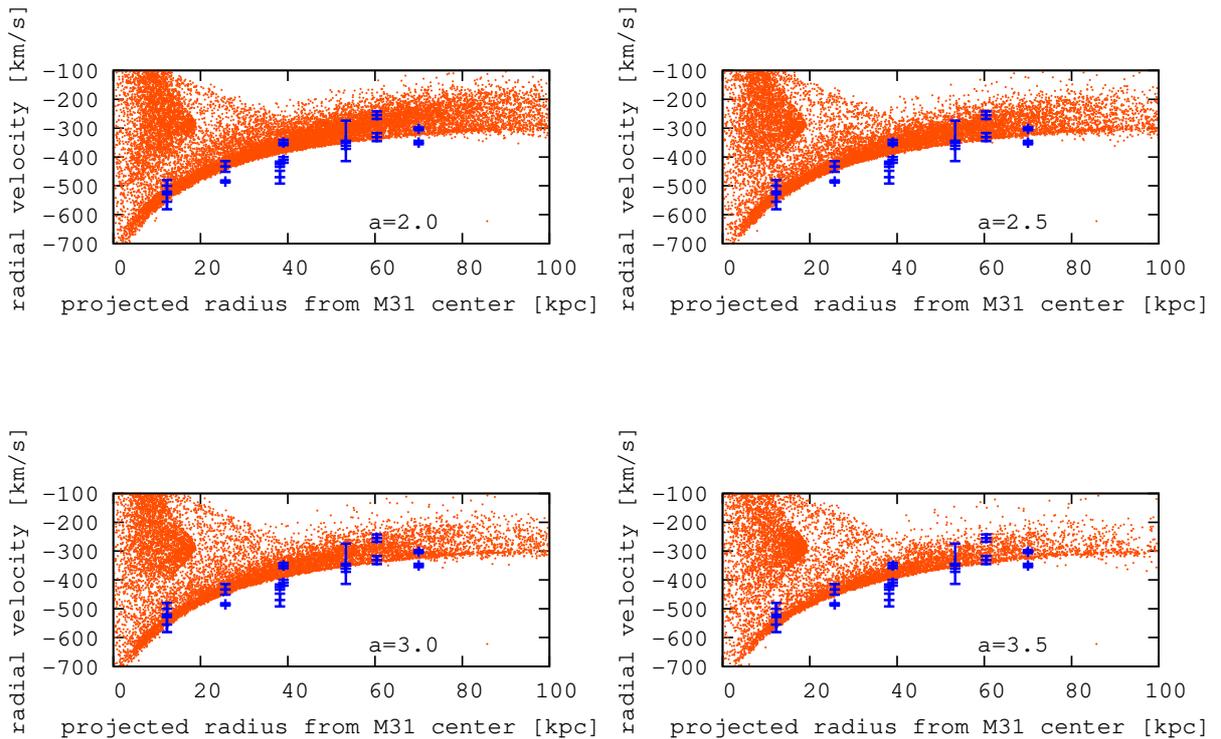}\hspace{2pc}%
 \end{center}
 \begin{minipage}[b]{38pc}\caption{\label{label2}Radial velocity distribution of the Andromeda giant stream.
Blue symbols show the observational data of \cite{r15}.
Red points show the results of $N$-body simulations in the stream region ($0^{\circ} \leq \xi \leq 8^{\circ}$ and $-10^{\circ} \leq \eta \leq 0^{\circ}$).
The upper-left panel shows the case of $a=2.0$, the upper-right panel shows the case of $a=2.5$, the lower-left panel shows the case of $a=3.0$ and the lower-right shows the case of $a=3.5$.}
 \end{minipage}
\end{figure}

Figure \ref{label2} shows the radial velocity distribution of the Andromeda giant stream.
The blue symbols in the figures are observed radial velocity distribution \cite{r15}. 
Red points show the results of $N$-body simulations.
It clearly shows that the radial velocity distribution of all simulations is consistent with that of observations.

\begin{figure}[h]
 \begin{center}
  \includegraphics[width=20pc]{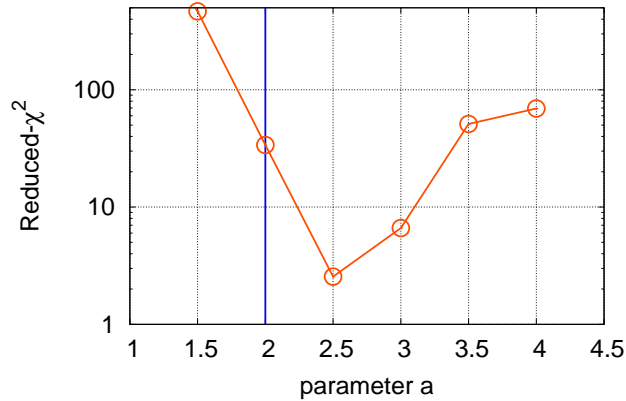}\hspace{2pc}%
 \end{center}
 \begin{minipage}[b]{38pc}\caption{\label{label3}Reduced-$\chi^2$ of the surface density ratio of the Andromeda giant stream, the eastern shell and the western shell as a function of density power-law slope $a$. The blue vertical line, $a=2$, is for the prediction by the cosmological $N$-body simulations based on the cold dark matter model.}
 \end{minipage}
\end{figure}

For the purpose of quantitative comparison between the observation and the simulation, we use the surface density ratio among the Andromeda giant stream and the two shells.
Figure \ref{label3} shows the result of $\chi^2$ analysis about the mutual combination of the density ratio among the Andromeda giant stream and the two shells.
Here, we show the levels of $\chi^2_{\nu}=2.3, 6.2$ and $11.8$, corresponding to $68, 95$ and $99.7$ percent confidence intervals in the case of three parameters.

We must note that the NFW model ($a=2$) fails to reproduce the apparent Andromeda giant stream and shell structures.
Only for the outer dark matter halo with a density power-law slope of $2 < a < 3.5$, the results of simulations successfully reproduce the observed Andromeda giant stream and shell structures.
The blue line, corresponds to the NFW model, which is the standard model suggested by cosmological $N$-body simulations. 

\section{Summary}
We examined the mass-density profile of the dark matter halo of M31 using $N$-body simulation of the galaxy collision.
As a result, the density profile of the dark matter halo suggested by the cosmological $N$-body simulation based on the CDM model fails to reproduce the Andromeda giant stream and the shell structures.
These structures are reproduced for $2 < a < 3.5$, and the best-fit parameter is $a=2.5$.
This result indicates that the mass-density profile of dark matter halo in M31 is steeper than that of the prediction of the CDM model. 
This discrepancy may be attributed to the triaxiality of the CDM halo, the tidal effect of M33 and Milky Way, and/or the morphology of the infalling satellite galaxy. Such models will be addressed in a forthcoming study.

\ack
We thank M. Umemura for use of the FIRST cluster at CCS, the University of Tsukuba.
This work was partially supported by the program of the Pre-Strategic Initiatives, University of Tsukuba, the FIRST project based on Grants-in-Aid for Specially Promoted Research by MEXT (16002003), and the Grant-in-Aid for Scientific Research (S)(2024002), (A)(21244013), and (C)(25400222).

\end{document}